\documentclass[prd,preprint,showpacs]{revtex4-1}
\usepackage[utf8]{inputenc}
\usepackage{amsmath}
\usepackage{latexsym}
\usepackage{amsfonts}
\usepackage{graphicx}
\usepackage{mathrsfs}
\usepackage{CJK}
\usepackage{longtable}

\usepackage{hyperref}
\hypersetup{
  colorlinks = true,
  urlcolor = blue,
  linkcolor = blue,
  citecolor = green,
  filecolor = magenta,
}

\newcommand{\ud}{\mathrm{d}}
\newcommand{\pd}{\partial}

\begin{document}
\begin{CJK*}{GB}{gbsn}

\title{Gravitational waves and the polarizations in Ho\v{r}ava gravity after GW170817}
\author{Yungui Gong}
\email{yggong@hust.edu.cn}
\affiliation{School of Physics, Huazhong University of Science and Technology, Wuhan, Hubei 430074, China}
\author{Shaoqi Hou}
\email{shou1397@hust.edu.cn}
\affiliation{School of Physics, Huazhong University of Science and Technology, Wuhan, Hubei 430074, China}
\author{Eleftherios Papantonopoulos}
\email{lpapa@central.ntua.gr}
\affiliation{Department of Physics, National Technical University of Athens, Zografou Campus GR 157 73, Athens, Greece}
\author{Dimitrios Tzortzis}
\email{dimitris.tzortzis@gmail.com}
\affiliation{Department of Physics, National Technical University of Athens, Zografou Campus GR 157 73, Athens, Greece}

\begin{abstract}
The gravitational waves of  Ho\v rava gravity, their  polarization states and their possible observational signatures   are discussed.
Using the gauge-invariant variable formalism, we find the three polarization modes in Ho\v rava gravity excited by the three physical degrees of freedom contained in this theory. In particular, the scalar degree of freedom excites a mix of the transverse breathing and the longitudinal polarizations.
The constraints from the previous experimental observations are taken into account, especially including the speed bound from the observations of GW170817 and GRB 170817A. It was found  that Ho\v rava theory is highly constrained.
Within the experimentally allowed parametric space, we studied whether the pulsar timing arrays and the Gaia mission can be used to distinguish the different polarizations.
After calculating the cross-correlation functions between the redshifts of photons and the astrometric positions of stars, one concludes that it is possible to tell whether there exits the scalar polarization using pulsar timing arrays and the Gaia mission.
\end{abstract}

\maketitle
\end{CJK*}

\section{Introduction}\label{sec-intro}

The LIGO Scientific and Virgo Collaborations have directly detected six gravitational wave (GW) events \cite{Abbott:2016blz,Abbott:2016nmj,Abbott:2017vtc,Abbott:2017oio,TheLIGOScientific:2017qsa,Abbott:2017gyy}.
Among them, the detection of GW170814   showed that the pure tensor polarizations are favored against pure vector and pure scalar polarizations \cite{Abbott:2017oio}.
GW170817 and GRB 170817A provided a very tight bound on the speed of GWs, and also heralded a new age of multimessenger astrophysics \cite{TheLIGOScientific:2017qsa,Goldstein:2017mmi,Savchenko:2017ffs}.
These events mark a new age when the nature of the gravity and General Relativity (GR) can be tested in the strong-field regime.
There are several different GW detectors at present and will be more in the future.
The ground-based interferometers, such as Advanced LIGO \cite{Harry:2010zz,TheLIGOScientific:2014jea}, Advanced Virgo \cite{TheVirgo:2014hva} and KAGRA \cite{Somiya:2011np,Aso:2013eba}, detect GWs in the high-frequency band (10$-10^4$ Hz).
These detectors will form a network in the coming year or so \cite{Aasi:2013wya}, providing a better way to probe the polarization content of GWs.
Instead, pulsar timing arrays (PTAs) \cite{Kramer:2013kea,Hobbs:2009yy,McLaughlin:2013ira,Hobbs:2013aka} detect GWs in the lower-frequency band (around $10^{-10}$$-10^{-6}$ Hz) \cite{Moore:2014lga}.
The Gaia mission and the alike are also capable of probing GWs within the similar frequency region \cite{Book:2010pf,2016A&A...595A...1G,Moore:2017ity,Klioner:2017asb,Mihaylov:2018uqm,OBeirne:2018slh}.
The space-borne interferometers are suitable to probe GWs in the intermediate-frequency band, such as LISA \cite{Audley:2017drz}, TianQin \cite{Luo:2015ght}, TaiJi \cite{Taiji2017}, DECIGO  \cite{Kawamura:2011zz} and so on.
In addition, atomic clocks are also sensitive to GWs in this frequency region \cite{Loeb:2015ffa,Kolkowitz:2016wyg,Graham:2017pmn}.

Alternative metric theories of gravity generally predict more GW polarization states than GR \cite{Will:2014kxa}.
By detecting the polarization content of GWs, one can test GR and its alternatives.
In this work, we will study the polarization content of a particular theory -- Ho\v{r}ava gravity \cite{Horava:2009uw}, and predict whether its polarization content can be detected by PTAs and Gaia mission \cite{2016A&A...595A...1G}.
Ho\v{r}ava gravity is a power-counting renormalizable theory of gravity.
The renormalizability is achieved by adding higher order spatial derivatives to the action.
This necessarily breaks the local Lorentz invariance.
In this theory, there is a preferred (3+1)-foliation, and in order to preserve this structure, the allowed diffeomorphisms are given by $t\rightarrow t'=T(t)$ and $x^j\rightarrow x'^j=X^j(t,x^k)$, which are called the foliation-preserving diffeomorphisms.
The breaking down of the local Lorentz invariance introduces one more degree of freedom (d.o.f.).
This new d.o.f. will excite a new GW polarization state.
The detection of the new polarization state will be the smoking gun indicating the departure from GR.
For a recent review on Ho\v rava gravity, please refer to Ref.~\cite{Wang:2017brl}.

The observations of GW170817 and GRB 170817A have questioned the validity of several alternatives theories of gravity.
Notably, as dark energy models, the Horndeski theory \cite{Horndeski:1974wa} and its generalizations \cite{Deffayet:2009mn} become simpler. That is, the functions of the scalar field $\phi$ and its kinetic energy $X=-\pd_\mu\phi\pd^\mu\phi/2$, $G_4(\phi,X)$ and $G_5(\phi,X)$ are severely constrained from the observations
  \cite{Baker:2017hug,Creminelli:2017sry,Sakstein:2017xjx,Ezquiaga:2017ekz,Langlois:2017dyl,Sakstein:2017xjx,Gong:2017kim}.
Constraints on alternative theories with vector fields were also considered.
For example, Ref.~\cite{Oost:2018tcv} considered the Einstein-\ae ther theory \cite{Jacobson:2000xp}, while Refs.~\cite{Gong:2018cgj,Hou:2018djz} discussed the bounds on both Einstein-\ae ther theory and generalized TeVeS theory \cite{Bekenstein:2004ne,Seifert:2007fr}.
Bimetric theories are also severely constrained.
It turns out that the two metrics should be proportional to each other if the matter fields couple with both of them (so these models are called the doubly coupled models), and the singly coupled models survive the speed bound \cite{Akrami:2018yjz}.

In the present work we will investigate the GW polarization states and their detection in Ho\v rava gravity by taking into account all the previous experimental constraints, including the recent GW speed bounds \cite{TheLIGOScientific:2017qsa,Goldstein:2017mmi,Savchenko:2017ffs}.
The GW solutions will be obtained  using the gauge-invariant variable formalism, and the polarization content is thus expressed in terms of the gauge-invariant variables.
The use of the gauge-invariant variables makes it easy to separate and identify the physical d.o.f., and the GW solutions can be obtained in an arbitrary gauge.
Ref.~\cite{Gumrukcuoglu:2017ijh} also discussed the constraints on this theory, but it did not address the problem of detecting the extra polarization states.

This work is organized in the following way.
In Section~\ref{sec-hggw}, we will briefly introduce Ho\v rava gravity and obtain its GW solutions about the Minkowski spacetime.
Then, we will identify the polarization content of GWs.
In Section~\ref{sec-cons}, we will discuss the previous experimental constraints.
Section~\ref{sec-exps} will be devoted to the investigation of the possibility to distinguish the different polarizations using PTAs and the Gaia mission.
Firstly, we will study the motions of stars and photons under the influence of the GW in Section~\ref{sec-msp}.
Next, we obtain the cross correlations of the redshifts of the photons coming from different pulsars for PTAs in Section~\ref{sec-pta}.
In Section~\ref{sec-gaia}, we calculate the cross correlations of the astrometric positions of distant stars for the Gaia mission.
Finally, the redshift and the astrometric position are also correlated, which will be computed in Section~\ref{sec-rac}.
Section~\ref{sec-con} is a brief summary.
Throughout this work, the geometrized units ($G=c=1$) will be used.

\section{Ho\v rava Gravity and its Gravitational Wave Solutions}\label{sec-hggw}

The low energy effective action of Ho\v{r}ava gravity can be conveniently expressed in terms of the Arnowitt-Deser-Misner variables \cite{Arnowitt:1962hi}
\begin{equation}\label{eq-act}
  S = \frac{1}{16\pi G_H}\int\ud^4xN\sqrt{g}[K_{jl}K^{jl}-(1+\lambda)K^2+(1+\beta)R+\alpha\nabla_j\ln N\nabla^j\ln N],
\end{equation}
where $G_H$ is the gravitational coupling constant, $N$ and $N_j$ are the lapse and shift functions, and $g$ is the determinant of the spatial metric tensor $g_{jl}$, so that the spacetime metric $g_{\mu\nu}$ is given by
\begin{equation}\label{eq-stmet}
  g_{\mu\nu}\ud x^\mu\ud x^\nu=-N^2\ud t^2+g_{jl}(\ud x^j+N^j\ud t)(\ud x^l+N^l\ud t).
\end{equation}
$R$ is the three dimensional Ricci scalar calculated using $g_{jl}$.
$K_{jl}=(\pd g_{jl}/\pd t-\nabla_jN_l-\nabla_lN_j)/2N$ is the extrinsic curvature tensor with $\nabla_j$ the covariant derivative compatible with $g_{jl}$.
There are three constants $\alpha$, $\beta$ and $\lambda$ which measure the differences from GR's action.
If  the action \eqref{eq-act} were required to be diffeomorphism invariant, $\alpha=\beta=\lambda=0$ \cite{Horava:2009uw} and then it would reduce to GR's.
Here, the action is only foliation-preserving diffeomorphism invariant, and these constants can be nonvanishing.

Ho\v{r}ava gravity at low energies can be viewed as a special case of Einstein-\ae{}ther theory with the \ae ther field satisfying the hypersurface orthogonal condition $u_{[\mu}\nabla_\nu u_{\rho]}=0$, where $\nabla_\mu$ is the covariant derivative compatible with the spacetime metric tensor $g_{\mu\nu}$ \cite{Blas:2010hb}.
Because of this extra constraint, there will be fewer degrees of freedom than in Einstein-\ae ther theory.
More specifically, there can be a scalar function $\phi$ connected to the \ae{}ther field through the relation
$u_\mu=-N\nabla_\mu\phi$ introducing in this way  only one extra d.o.f. as it will be discussed below.
$\phi$ is also called the ``khronon", and the action \eqref{eq-act} is that of the khronometric theory \cite{Germani:2009yt,Blas:2010hb}.

Varying this action with respect to  $N$, $N_j$ and $g^{jl}$, one obtains the following equations of motion,
\begin{gather}
  \frac{1}{16\pi G_H}[(1+\beta)R-K_{jl}K^{jl}+(1+\lambda)K^2-\alpha\nabla_j\ln N\nabla^j\ln N
  -2\alpha\nabla_j\nabla^j\ln N] = 0, \\
  \frac{1}{8\pi G_H}[\nabla_lK^{lj}-(1+\lambda)\nabla^jK] = 0, \\
  \frac{1}{16\pi G_H}\Big\{(1+\beta)\Big(R_{jl}-\frac{1}{2}g_{jl}R\Big)+2K_{jk}K^k_l-\frac{1}{2}g_{jl}K_{km}K^{km}
  \nonumber\\
  -2(1+\lambda)(KK_{jl}-\frac{1}{4}g_{jl}K^2)+\alpha\nabla_j\ln N\nabla_l\ln N-\alpha\frac{1}{2}g_{jl}(\nabla_k\ln N)\nabla^k\ln N
  \nonumber\\
  +\frac{g_{jk}g_{lm}}{N\sqrt{g}}\pd_t(\sqrt{g}K^{km})-(1+\lambda)\frac{g_{jk}g_{lm}}{N\sqrt{g}}\pd_t(\sqrt{g}Kg^{km})
  \nonumber\\
  +\frac{1+\beta}{N}(g_{jl}\nabla_k\nabla^k N-\nabla_j\nabla_lN)+\frac{1}{N}\nabla_k(2N_{(j}K_{l)}^k-N^kK_{jl})
  \nonumber\\
  -\frac{1+\lambda}{N}[2\nabla_{(j}(N_{l)}K)-g_{jl}\nabla_k(N^kK)]\Big\}= 0.
\end{gather}
After linearization, these equations are greatly simplified.
Further simplification can be achieved by using that fact that
the action \eqref{eq-act} is invariant under the foliation-preserving diffeomorphisms.
Infinitesimally, the gauge transformation is \cite{Horava:2009uw}
\begin{gather}\label{eq-gt}
  \delta N=\xi^k\pd_kN+\dot N\xi_0+N\dot\xi_0, \\
  \delta N_j=N_k\pd_j\xi^k+\xi^k\pd_kN_j+g_{jk}\dot\xi^k+\dot N_j\xi_0+N_j\dot\xi_0, \\
  \delta g_{jk}=\xi^l\pd_lg_{jk}+g_{jl}\pd_k\xi^l+g_{lk}\pd_k\xi^j+\xi_0\dot g_{jk},
\end{gather}
generated by $\xi_\mu=(\xi_0,\xi_j)$ with $\xi_0=\xi_0(t)$.
These transformation laws can be obtained by taking $c\rightarrow\infty$ limit of the usual transformation laws for the spacetime metric tensor $g_{\mu\nu}$ \cite{Horava:2009uw}.
Now, let us determine the GW solutions about the Minkowski spacetime with $N=1,N_j=0$, and $g_{jk}=\delta_{jk}$.
Assume the perturbed spacetime metric is given by
\begin{equation}\label{eq-pstm}
  N=1+n,\quad N_j=n_j,\quad g_{jk}=\delta_{jk}+h_{jk}.
\end{equation}
Under the gauge transformation of Eq.~\eqref{eq-gt}, one knows that
\begin{gather*}
  n\rightarrow n+\dot\xi_0,\quad n_j\rightarrow n_j+\dot\xi_j,\\ h_{jk}\rightarrow h_{jk}+\pd_j\xi_k+\pd_k\xi_j.
\end{gather*}
Now, decompose $n_j$ and $h_{jk}$ into their transverse and longitudinal parts as in the following way,
\begin{gather}
n_j=\beta_j+\pd_j\gamma,\label{eq-nj-dec}\\
h_{jk} = h_{jk}^\mathrm{TT}+\frac{1}{3}H\delta_{jk}+\partial_{(j}\epsilon_{k)}+\left(\partial_j\partial_k-\frac{1}{3}\delta_{jk}\nabla^2\right)\rho.\label{eq-hjk-dec}
\end{gather}
In these expressions, $\gamma$, $H$ and $\rho$ are scalars.
$H=\delta^{jk}h_{jk}$ is the trace.
$\beta_j$ and $\epsilon_j$ are transverse vectors, and $h_{jk}^\text{TT}$ is the transverse-traceless part of $h_{jk}$.
They satisfy $\pd^j\beta_j=\pd^j\epsilon_j=\delta^{jk}h_{jk}^\text{TT}=0$ and $\pd^k h_{jk}^\text{TT}=0$.
Then one can define the following ``restricted" gauge-invariant variables
\begin{gather}\label{eq-ginv}
  n,\quad h_{jk}^\text{TT},\\
  \Xi_j=\beta_j-\frac{1}{2}\dot\epsilon_j,\\
  \Theta=\frac{1}{2}(H-\nabla^2\rho),\\
  \Psi=\gamma-\frac{1}{2}\dot\rho.
\end{gather}
These variables are invariant under the restricted transformation with $\xi_0=0$.
Then, the linearized equations of motion lead to
\begin{gather}
  \frac{3\lambda+2}{\lambda}\ddot\Theta-\frac{(1+\beta)(2+2\beta-\alpha)}{\alpha}\nabla^2\Theta=0, \label{eq-eom-Theta}\\
  \ddot h_{jk}^\text{TT}-(1+\beta)\nabla^2h_{jk}^\text{TT}=0,\label{eq-eom-spin2} \\
  \Xi_j=0,\quad n=-\frac{1+\beta}{\alpha}\Theta,\quad\nabla^2\Psi=\frac{3\lambda+2}{2\lambda}\dot\Theta.\label{eq-eom-aux}
\end{gather}
Therefore, there are three propagating degrees of freedom represented by $h_{jk}^\text{TT}$ and $\Theta$.
The squared speeds  can be easily read off, given by
\begin{gather}
  s_2^2=1+\beta,\label{eq-s2sq}\\
   s_0^2=\frac{\lambda(1+\beta)(2+2\beta-\alpha)}{\alpha(3\lambda+2)},\label{eq-s0sq}
\end{gather}
for the tensor and scalar GWs, respectively.
When both speeds are 1, one has the following conditions,
\begin{equation}\label{eq-spds-1}
  \beta=0,\quad\alpha+(2\alpha-1)\lambda=0.
\end{equation}

Since the gravity only enjoys the foliation-preserving diffeomorphism in Ho\v{r}ava gravity, the matter action could take a form that is different from the one in GR \cite{Capasso:2009fh,Kimpton:2013zb}.
For example, a simple action for the point particle of mass $m$ is $-m\int\ud\tau \sqrt{g_{\mu\nu}\dot x^\mu\dot x^\nu+\sigma(N\dot t)^2}$, where $\dot x^\mu=\ud x^\mu/\ud\tau$ with $\tau$ the proper time, and $\sigma$ is a coupling constant.
However, this action and the alike  violate the local Lorentz symmetry in the matter sector, which has been severely constrained \cite{Mattingly:2005re}.
So we will assume that the matter fields minimally couple with the spacetime metric $g_{\mu\nu}$ as in GR.

If matter fields couple with the spacetime metric minimally, test particles will follow geodesics determined by $g_{\mu\nu}$.
In order to determine the polarizations, one needs to calculate the linearized geodesic deviation equations, $\ddot x^j=d^2x^j/dt^2=-\mathcal R_{tjtk}x^k$ with $\mathcal R_{tjtk}$ calculated using $g_{\mu\nu}$.
The electric component $\mathcal R_{tjtk}$ is \cite{Flanagan:2005yc}
\begin{equation}\label{eq-rtjtk}
 \mathcal R_{tjtk}=-\frac{1}{2}\ddot h_{jk}^\mathrm{TT}+\dot\Xi_{(j,k)}+\Phi_{,jk}-\frac{1}{2}\ddot\Theta\delta_{jk},
\end{equation}
where in this case, $\Phi=-n/2+\dot\gamma-\ddot\rho/2=\frac{(1+\beta)(3+2\beta-\alpha)}{2\alpha}\Theta$.
Therefore,
\begin{equation}\label{eq-rtjtk-2}
\mathcal R_{tjtk}=-\frac{1}{2}\ddot h_{jk}^\mathrm{TT}+\frac{(1+\beta)(3+2\beta-\alpha)}{2\alpha}\pd_j\pd_k\Theta-\frac{1}{2}\delta_{jk}\ddot\Theta.
\end{equation}

To extract the polarization content explicitly, one assumes that the GW propagates in the $+z$ direction with the following wave vectors
\begin{gather}\label{eq-wvs}
  k_2^\mu=\omega_2(1,0,0,1/s_2),\\
   k_0=\omega_0(1,0,0,1/s_0),
\end{gather}
for the tensor and scalar GWs, respectively, and $\omega_2$ and $\omega_0$ are the corresponding angular frequencies.
One finds that the nonvanishing components of the 4-dimensional Riemann tensor $\mathcal R_{tjtk}$ are
\begin{gather}
 \mathcal R_{txtx}=\frac{\omega_2^2}{2}h_{xx}^\text{TT}+\frac{\omega_0^2}{2}\Theta,\label{eq-rtxtx} \\
 \mathcal R_{txty}=\frac{\omega_2^2}{2}h_{xy}^\text{TT},\label{eq-rtxty} \\
 \mathcal R_{tyty}=-\frac{\omega_2^2}{2}h_{xx}^\text{TT}+\frac{\omega_0^2}{2}\Theta,\label{eq-rtyty}  \\
 \mathcal R_{tztz}=-\frac{\omega_0^2}{2}\left(\frac{3\lambda+2}{\lambda}\frac{3+2\beta-\alpha}{2+2\beta-\alpha}-1\right)\Theta.\label{eq-rtztz}
\end{gather}
One immediately recognizes the $+$ and $\times$ polarizations as in GR excited by $h_{xx}^\text{TT}=-h_{yy}^\text{TT}=h_+$ and $h_{xy}^\text{TT}=h_{yx}^\text{TT}=h_\times$, respectively.
The scalar degree of freedom $\Theta$ excites the transverse breathing polarization as in scalar-tensor theory \cite{Hou:2017bqj,Gong:2017bru}, Einstein-\ae ther theory and generalized TeVeS theory \cite{Gong:2018cgj,Hou:2018djz,Gong:2018ybk}.
$\Theta$ also excites the longitudinal polarization as long as $\mathcal R_{tztz}\ne0$.
When the following condition
\begin{equation}\label{eq-rtztz-0}
  \lambda=-2\frac{3+2\beta-\alpha}{7+2(2\beta-\alpha)}
\end{equation}
is satisfied, the longitudinal polarization disappears, i.e., $\mathcal R_{tztz}=0$.

Therefore, one concludes that there are three polarization states: the $+$ and $\times$ polarization states excited by $h_+$ and $h_\times$, respectively, and the mixed state of the breathing and longitudinal polarizations excited by the scalar field $\Theta$.
In the next sections, we will first review the previous experimental constraints on Ho\v{r}ava gravity, and then discuss whether it is possible to distinguish the different polarizations with PTAs and Gaia mission.

\section{Experimental Constraints}\label{sec-cons}

Since the proposal of Ho\v{r}ava gravity, there have been several theoretical and experimental constraints as collected in
Ref.~\cite{Gumrukcuoglu:2017ijh}.
The first three of them are \cite{Blas:2009qj}
\begin{enumerate}
  \item Unitarity: $\lambda(3\lambda+2)>0$,
  \item Perturbative stability: $0<\alpha/(1+\beta)<2$ and $\beta/(1+\beta)<1$,
  \item Big bang nucleosynthesis: $|(\alpha-2\beta+3\lambda)/(3\lambda+2)|<1/8$.
\end{enumerate}
We will also require that the GW speeds should be at least 1 in order to forbid the gravitational Cherenkov radiation \cite{Elliott:2005va,Jacobson:2008aj},
\begin{equation}\label{eq-cons-gcr}
  s_2\ge1,\quad s_0\ge1,
\end{equation}
which was also demanded in Ref.~\cite{Gumrukcuoglu:2017ijh}.
Then the post-Newtonian parameters $\alpha_1$ and $\alpha_2$ are highly bounded  \cite{Blas:2011zd,Freire:2012mg,Shao:2013wga,Shapiro:1999fn},
\begin{gather}
  |\alpha_1|=|-4\alpha+8\beta|\le4\times10^{-5},\label{eq-cons-alpha-1}\\
  |\alpha_2|=\left|\frac{(\alpha-2\beta)[\lambda-(\alpha-2\beta)(1+2\lambda)]}{\lambda(\alpha-2\beta-2)}\right|\le2\times10^{-9}.\label{eq-cons-alpha-2}
\end{gather}
Finally, the tensor speed $s_2$ is constrained by the recent observations on GW170817 and GRB 170817A so that \cite{TheLIGOScientific:2017qsa,Monitor:2017mdv}
\begin{equation}\label{eq-cons-s2}
  -3\times10^{-15}\le s_2-1\le7\times10^{-16}.
\end{equation}

Constraints based on binary pulsar observations were obtained in Ref.~\cite{Yagi:2013ava}.
They were given in Figures 1 (the right panel) and 8 with the restriction that $\alpha=2\beta$.
This restriction came from the fact that Eqs.~\eqref{eq-cons-alpha-1} and \eqref{eq-cons-alpha-2} show that $\alpha_1$ and $\alpha_2$ are highly constrained, so authors of Ref.~\cite{Yagi:2013ava} expanded the theory in powers of $\alpha_1$ and $\alpha_2$, and considered the part at the leading order ($\alpha_1=\alpha_2=0$ implies $\alpha=2\beta$).
This choice was sufficient before the constraint on the GW speed \eqref{eq-cons-s2}.
With the advent of the new constraint \eqref{eq-cons-s2}, it is better to abandon  the restriction $\alpha=2\beta$, following Ref.~\cite{Gumrukcuoglu:2017ijh}.
If this relation is not valid, it is possible that a large scalar speed  increases the decay of a binary system and this is not happening in the opposite case.

Taking all the constraints into account, one obtains that
\begin{gather}
  0\le\beta\le1.4\times10^{-15},\label{eq-cons-beta}\\
  0<\lambda\lesssim\lambda_1=0.0952381.\label{eq-cons-alpha}
\end{gather}
The constraint on $\alpha$ is more complicated and depends on $\beta$ and $\lambda$.
Since $\beta$ is severely bounded from above, one can consider two special cases.
In the first case, let $\beta=0$,
so the tensor GW propagates at the speed of light, $s_2=1$.
The bounds on $\alpha$ are given by
\begin{equation}\label{eq-cons-alpha-beta-0}
  0<\alpha\le\left\{
  \begin{array}{cc}
    \displaystyle\frac{\lambda}{1+2\lambda}, & 0<\lambda\lesssim10^{-4}, \\
    10^{-5}, &  10^{-4}\lesssim\lambda\lesssim\lambda_2,\\
    \displaystyle\frac{2-21\lambda}{8}, & \lambda_2\lesssim\lambda\lesssim\lambda_1,
  \end{array}\right.
\end{equation}
where $\lambda_2=0.0952343$.
In the second case, set $\beta=1.4\times10^{-15}$. Then the lower bound on $\alpha$ is still 0, and the upper bound changes a little.
From the above analysis, one finds out that the parameters $\alpha$ and $\beta$ are highly constrained.
To achieve such small values, a severe fine tuning is required.

For purpose of making definite predictions in the next section, let us choose some points in the allowed parameter space so that $s_2=1$ and the choices are listed in Table~\ref{tab-choices-beta-0}.
These choices make sure that the scalar GW propagates at the superluminal speeds $s_0$.
In addition, these speeds are required not to be too large.
This is because a very large speed $s_0$ might lead to a faster decay of the binary star system, as the scalar GW would carry away energy faster.
\begin{table}
  \centering
    \caption{Some choices for the values for $\alpha$ and $\lambda$ at $\beta=0$ and the speed $s_0$.
    Note that $\alpha$ and $\lambda$ are normalized by $10^{-7}$.}\label{tab-choices-beta-0}
  \begin{tabular}{ccccc}
    \hline\hline
    $\alpha$ & &$\lambda$ & &$s_0$ \\
    \hline
    100 &$\quad$ & 100.2&$\quad$ & 1.001 \\
    10 & $\quad$& 13.33 &$\quad$ & 1.155 \\
    1 & $\quad$& 3.00 & $\quad$&1.732 \\
    \hline
  \end{tabular}
\end{table}

\section{Experimental Tests}\label{sec-exps}

When the GW passes by, the motions of stars and photons will be affected.
Firstly, the propagation time of photons emitted from pulsars or stars to the Earth changes, which can be measured by PTAs \cite{Lee2008ptac}.
Secondly, the apparent positions, i.e., the astrometric positions, of stars in the sky also change due to the deflected trajectories of photons.
The change in the astrometric positions is monitored by the Gaia mission launched in 2013 by the European Space Agency \cite{2016A&A...595A...1G}.
Both  projects can detect the polarizations of GWs \cite{Hellings:1983fr,Lee2008ptac,Book:2010pf,Moore:2017ity,Klioner:2017asb,Mihaylov:2018uqm,OBeirne:2018slh}.
For this purpose, one needs to first study the motions of stars and photons affected by the monochromatic plane GWs, and then by the stochastic GW background in Ho\v{r}ava gravity.

\subsection{The motion of stars and photons}\label{sec-msp}

For this end, one has to fix the ``restricted" gauge by setting $n_j=0$, so $\beta_j=0$ and $\gamma=0$.
Since $\Xi_j=\beta_j-\dot\epsilon_j/2=0$, $\epsilon_j=0$ up to a function of position only.
One also obtains that
\begin{equation}\label{eq-fg-r}
  \Theta=-\frac{\lambda}{\lambda+2}H,\quad n=-\frac{1+\beta}{\alpha}\Theta,\quad\nabla^2\rho=-\frac{3\lambda+2}{\lambda}\Theta.
\end{equation}
Let the monochromatic plane waves be described by
\begin{gather}\label{eq-mono-w}
  h_{jk}^\text{TT}(t,\vec x)=A_{jk}^\text{TT}\cos\left[\omega_2\left(t-\hat\Omega\cdot \vec x/s_2\right)\right],\\
  \Theta(t,\vec x)=\Theta_0\cos\left[\omega_0\left(t-\hat\Omega\cdot\vec x/s_0\right)\right],
\end{gather}
where $A_{jk}^\text{TT}$ and $\Theta_0$ are the amplitudes, and the GW is propagating in the direction $\hat\Omega$.
Note that these GWs are propagating at speeds other than 1, which is different from the analyses done in Refs.~\cite{Hellings:1983fr,Lee2008ptac,Book:2010pf,Moore:2017ity,Klioner:2017asb,Mihaylov:2018uqm,OBeirne:2018slh}.

First, let us study the motion of a massive particle, modeling the Earth or the star, influenced by GWs.
It is assumed that in the absence of GWs, the Earth is at the origin of the coordinate system, and the star is at $\vec x_\star^{(0)}=L\hat r$ with $L$ the distance from the Earth to the star.
When the GW passes by, it might acquire a nontrivial 4-velocity $u^\mu_p=u^0_p(1,\vec v_p)$ where $p$ stands for the Earth ($\oplus$) or a star ($\star$).
Calculation shows that the 4-velocities of the Earth and the star are given by
\begin{gather}
  u^\mu_\oplus=\delta^\mu_0+\frac{1+\beta}{\alpha}\left(1,-\frac{\hat\Omega_j}{s_0}\right)\Theta(t,0),\label{eq-e4v} \\
  u^\mu_\star=\delta^\mu_0+\frac{1+\beta}{\alpha}\left(1,-\frac{\hat\Omega_j}{s_0}\right)\Theta(t,L\hat r),\label{eq-s4v}
\end{gather}
respectively.
Using these results, one can easily determine the trajectories of the Earth and the star
\begin{gather}
  x^j_\oplus(t)=-\frac{\hat\Omega^j}{\omega_0s_0}\frac{1+\beta}{\alpha}\Theta_0\sin\omega_0t, \label{eq-x-star}\\
  x^j_\star(t)=L\hat r^j-\frac{\hat\Omega^j}{\omega_0s_0}\frac{1+\beta}{\alpha}\Theta_0\sin\omega_0\left(t-L\hat\Omega\cdot\hat r/s_0\right),\label{eq-x-e}
\end{gather}
so that when the GW is absent, the Earth is at the origin and the star is at $L\hat r$.
This means that only the scalar GW affects the motions of the Earth and the star.

In order to calculate the change in the astrometric position of the star, one first chooses a tetrad basis $e^\mu_{\hat a}$ ($a=0,1,2,3$) which defines the proper reference frame of an observer comoving with the Earth \cite{Mihaylov:2018uqm}.
So $e^\mu_{\hat 0}=u^\mu_\oplus$, and the tetrads are parallel transported along the geodesic of the observer.
Since we have already obtained $e^\mu_{\hat 0}$, we only need determine the triad $e^\mu_{\hat j}$ ($\hat j=1,2,3$).
In the absence of the GW, it is natural to choose $\bar e^\mu_{\hat j}=\delta^\mu_{\hat j}$, and the GW induces perturbations so that $e^\mu_{\hat j}=\bar e^\mu_{\hat j}+\epsilon^\mu_{\hat j}$ with $\epsilon^\mu_{\hat j}$ of the same order as the metric perturbation.
The parallel transport equations for $e^\mu_{\hat j}$ are
\begin{equation}\label{eq-patr}
\begin{split}
  0&=u^\nu_\oplus\pd_\nu e^\mu_{\hat j}+\Gamma^\mu{}_{\rho\nu}u^\nu_\oplus e^\rho_{\hat j}\\
  &\approx\frac{\ud}{\ud t} \epsilon^\mu_{\hat j}+\Gamma^\mu{}_{0\hat j},
\end{split}
\end{equation}
which are approximately evaluated along the zeroth order worldline of the Earth, i.e., the time axis.
The calculation results in
\begin{equation}
\epsilon^\mu_{\hat j}=\left\{
\begin{array}{cc}
 \displaystyle -\frac{\hat\Omega_j}{s_0}\frac{1+\beta}{\alpha}\Theta(t,0), & \mu=0, \\
\displaystyle -\frac{1}{2}h_{jk}^\text{TT}(t,0)-\left(\frac{1}{3}\delta_{jk}-\frac{3\lambda+2}{2\lambda}\hat\Omega_j\hat\Omega_k\right) \Theta(t,0), & \mu=k.
\end{array}\right.
\end{equation}

Next, we study how the photon trajectories are affected by GWs.
Let us assume that in the absence of the GW, the photon's 4-velocity is $\bar u_\gamma^\mu=\gamma_0(1,-\hat r)$, and the GW perturbs its 4-velocity so that $u^\mu_\gamma=\bar u^\mu_\gamma+V^\mu$.
After some tedious calculation, one obtains the following results,
\begin{gather}
  V^0=-\gamma_0\left\{\frac{\hat r^j\hat r^kh_{jk}^\text{TT}}{2(1+\hat \Omega\cdot\hat r/s_2)}+\frac{\Theta}{1+\hat\Omega\cdot\hat r/s_0}\Bigg[\frac{1}{3}-\frac{1+\beta}{\alpha}\left(1+2\frac{\hat\Omega\cdot\hat r}{s_0}\right)\right. \nonumber\\
  \left.-\frac{3\lambda+2}{2\lambda}(\hat\Omega\cdot\hat r)^2\Bigg]+\nu^0\right\},\\
  V^j=-\gamma_0\left\{-h_{jk}^\text{TT}\hat r^k+\frac{\hat \Omega^j\hat r^k\hat r^lh_{kl}^\text{TT}}{2(s_2+\hat\Omega\cdot\hat r)}-\frac{2}{3}\hat r^j\Theta+\Bigg[\frac{1}{3}+\frac{1+\beta}{\alpha}+\frac{3\lambda+2}{2\lambda}\hat\Omega\cdot\hat r\right.\nonumber\\
 \left. (2s_0+\hat\Omega\cdot\hat r)\Bigg]\frac{\hat\Omega^j\Theta}{s_0+\hat\Omega\cdot\hat r}+\nu^j\right\},
\end{gather}
where $\nu^0$ and $\nu^j$ are integration constants and of the same order as the perturbations, and $h_{jk}^\text{TT}$ and $\Theta$ are both evaluated at $(t,0)$.
The condition that $g_{\mu\nu}u^\mu_\gamma u^\nu_\gamma=0$ translates to $\nu^0+\hat r^j\nu_j=0$.

The photon trajectory can thus be determined,
\begin{equation}\label{eq-gam-traj}
\begin{split}
  x^j_\gamma(t)=&-\Bigg\{(t-L-t_e)\hat r^j+(\nu^0\hat r^j+\nu^j)t+x_0^j-\frac{A_{jk}^\text{TT}\hat r^k\sin\Phi_2(t)}{\omega_2(1+\hat\Omega\cdot\hat r/s_2)}\\
  &+\frac{\hat r^j+\hat\Omega^j/s_2}{2\omega_2(1+\hat\Omega\cdot\hat r/s_2)^2}\hat r^k\hat r^lA_{kl}^\text{TT}\sin\Phi_2(t)+\Big[\frac{1}{3}-\frac{1+\beta}{\alpha}\Big(1+2\frac{\hat\Omega\cdot\hat r}{s_0}\Big)\\
  &-\frac{3\lambda+2}{2\lambda}(\hat\Omega\cdot\hat r)^2\Big]\frac{\hat r^j\Theta_0\sin\Phi_2(t)}{\omega_0(1+\hat\Omega\cdot\hat r/s_0)^2}+\Big[\frac{1}{3}+\frac{1+\beta}{\alpha}+\frac{3\lambda+2}{2\lambda}\times\\
  &\hat\Omega\cdot\hat r(2s_0+\hat\Omega\cdot\hat r)\Big]\frac{s_0\hat\Omega^j\Theta_0\sin\Phi_0(t)}{\omega_0(s_0+\hat\Omega\cdot\hat r)^2}\Bigg\},
  \end{split}
\end{equation}
where $x_0^j$ is the integration constant of the same order as the perturbations, and the phases are $\Phi_2(t)=\omega_2[t-(L+t_e-t)\hat\Omega\cdot\hat r/s_2]$ and $\Phi_0(t)=\omega_0[t-(L+t_e-t)\hat\Omega\cdot\hat r/s_0]$.
Eqs.~\eqref{eq-x-star}, \eqref{eq-x-e} and \eqref{eq-gam-traj} all depend on the parameters $\alpha$, $\beta$ and $\lambda$ as expected.
In contrast, the photon trajectory \eqref{eq-gam-traj} depends on both the scalar and the tensor modes.
Note also that all trajectories \eqref{eq-x-star}, \eqref{eq-x-e} and \eqref{eq-gam-traj} are related to the speeds $s_0$ and $s_2$ of the scalar and the tensor modes.

\subsection{Pulsar timing arrays}\label{sec-pta}

Now, one can calculate the frequency shift.
The frequencies of the photon measured by the observers comoving with the Earth and the star are $f_\oplus=-u^\mu_\gamma u^\oplus_\mu$ and $f_\star=-u^\mu_\gamma u^\star_\mu$, respectively.
Then the relative frequency shift is
\begin{equation}\label{eq-fs}
\begin{split}
  \frac{f_\star-f_\oplus}{f_\oplus}=&I_0(\hat\Omega,\hat r)[\Theta(t,0)-\Theta(t-L,L\hat r)]\\
  &+I_2^{jk}(\hat\Omega,\hat r)[h_{jk}^\text{TT}(t,0)-h_{jk}^\text{TT}(t-L,L\hat r)],
\end{split}
\end{equation}
with
\begin{gather}
  I_0(\hat \Omega,\hat r)=\frac{1}{1+\hat\Omega\cdot\hat r/s_0}\Bigg[\frac{1}{3}-\frac{3\lambda+2}{2\lambda}(\hat\Omega\cdot\hat r)^2-\frac{1+\beta}{\alpha}\Big(\frac{\hat\Omega\cdot\hat r}{s_0}\Big)^2\Bigg],\label{eq-def-i0}\\
  I_2^{jk}(\hat \Omega,\hat r)=\frac{\hat r^j\hat r^k}{2(1+\hat \Omega\cdot \hat r/s_0)}.\label{eq-def-i2}
\end{gather}

The stochastic GW background can be described by
\begin{gather}
  \Theta(t,\vec x)=\int_{-\infty}^{\infty}\frac{\ud \omega }{2\pi}\int\ud^2\hat\Omega \Big\{\Theta(\omega ,\hat\Omega )\exp[i(\omega t-k \hat\Omega \cdot\vec x)]\Big\},\label{eq-sgw-scl}\\
  h_{jk}^\text{TT}(t,\vec x)=\sum_{P=+,\times}\int_{-\infty}^{\infty}\frac{\ud\omega}{2\pi}\int\ud^2\hat\Omega\Big\{\epsilon_{jk}^Ph_P(\omega,\hat\Omega)\exp[i(\omega t-k \hat\Omega \cdot\vec x)]\Big\}, \label{eq-tn-ex}
\end{gather}
where $\Theta(\omega, \hat k)$ and $h_P(\omega,\hat k)$ are the amplitudes of the scalar and tensor GWs oscillating at $\omega$ and propagating in the direction  $\hat k$, respectively.
$\epsilon^P_{jk}$ is the polarization matrix and $P=+,\times$.
Suppose that the stochastic GW background is isotropic, stationary, and independently polarized; then, one defines the characteristic strains
 $\Theta_c(\omega)$ and $h_c^P(\omega )$ in the following manner
\begin{gather}
\langle \Theta^*(\omega ,\hat\Omega )\Theta(\omega' ,\hat\Omega' )\rangle=\delta(\omega -\omega' )\delta^{(2)}(\hat\Omega -\hat\Omega' )\frac{|\Theta_c(\omega )|^2}{\omega },\label{defvc}\\
  \langle h^{*}_P(\omega ,\hat\Omega)h_P(\omega ',\hat\Omega')\rangle=\delta(\omega -\omega ')\delta^{(2)}(\hat\Omega-\hat\Omega')\delta^{PP'}\frac{\pi|h_c^P(\omega )|^2}{4\omega},  \label{eq-def-s-h}
\end{gather}
where a star $*$ indicates  complex conjugation.
The characteristic strains are proportional to $\omega^\alpha$ with $\alpha$ called the power-law index.

Integrating the relative frequency shift gives the timing residual \cite{Lee2008ptac}
\begin{equation}\label{rtsl}
  R(T)=\int_{-\infty}^{\infty}\frac{\ud \omega }{2\pi}\int\ud^2\hat \Omega \int_{0}^{T}\ud t\frac{f_\star-f_\oplus}{f_\oplus},
\end{equation}
where the argument $T$ is the total observation time.
The cross-correlation function $C(\theta)=\langle R_a(T)R_b(T)\rangle$ can thus be obtained.

\begin{figure}
  \centering
  \includegraphics[width=0.5\textwidth]{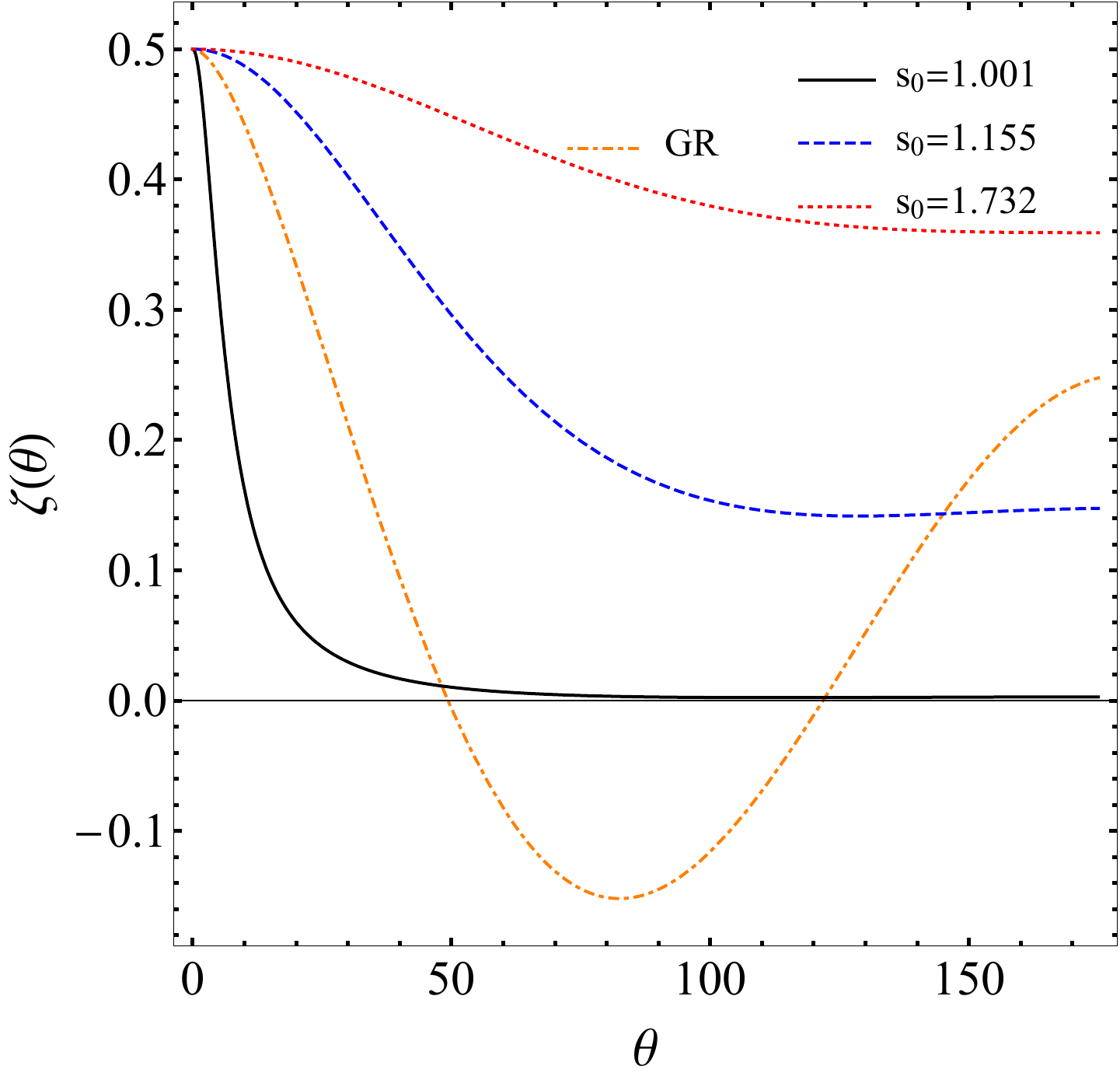}
  \caption{The normalized cross-correlation function $\zeta(\theta)$ for the scalar polarization at different speeds.
  Also shown is the one for the tensor polarizations in GR (the dot-dashed orange curve, labeled by ``GR"), which also represents the normalized cross-correlation function for the tensor polarizations for Ho\v rava gravity.}\label{fig-scl-xcor}
\end{figure}

Now, we can calculate the cross correlation function for the scalar and tensor GWs in sequence.
For the scalar GW, the timing residual is
\begin{equation}\label{eq-tr-scl}
  R(T)=\int_{-\infty}^{\infty}\frac{\ud\omega}{2\pi}\int\ud^2\hat\Omega\left\{I_0(\hat\Omega,\hat r)\Theta(\omega,\hat\Omega)\frac{e^{i\omega T}-1}{i\omega}\left[1-e^{-i\omega L(1+\hat\Omega\cdot\hat r/s_0)}\right]\right\}.
\end{equation}
So the cross correlation $C_0(\theta)$ between two pulsars $a$ and $b$ located at $\vec x_1=L_1\hat r_1$ and $\vec x_2=L_2\hat r_2$, respectively, is
\begin{equation}\label{eq-xc-scl}
  C_0(\theta)=\int_{0}^{\infty}\frac{\ud\omega}{2\pi^2}\int\ud^2\hat\Omega\frac{|\Theta_c(\omega)|^2}{\omega^3}I_0(\hat\Omega,\hat r_1)I_0(\hat\Omega,\hat r_2)\mathcal P_0,
\end{equation}
where $\mathcal P_0=1-\cos\Delta_1-\cos\Delta_2+\cos(\Delta_1-\Delta_2)$ with $\Delta_j=\omega L_j(1+\hat k\cdot\hat r_j/s_0)\;(j=1,2)$.
To get this result, one also averages over $T$, as required by the ensemble average \cite{Lee2008ptac}.
Now, following the similar argument made in Ref.~\cite{Gong:2018cgj}, one can easily calculate $C_0(\theta)$ and obtain the so-called normalized cross-correlation function $\zeta(\theta)=C_0(\theta)/C_0(0)$, which is shown in Fig.~\ref{fig-scl-xcor}.
This figure shows $\zeta(\theta)$ at different values of the speed $s_0$ listed in Table~\ref{tab-choices-beta-0}.
For the tensor GW, since its speed $s_2=1$, the normalized cross-correlation function has been obtained \cite{Hellings:1983fr},
\begin{equation}\label{eq-xcor-hd}
  \zeta_2(\theta)=\frac{3}{4}(1-\cos\theta)\ln\frac{1-\cos\theta}{2}-\frac{1-\cos\theta}{8}+\frac{1+\delta(\theta)}{2}.
\end{equation}
$\zeta_2(\theta)$ is also plotted in Fig.~\ref{fig-scl-xcor}, represented by the dot-dashed orange curve.
From this figure, one can clearly see that the tensor polarizations induce a very different cross correlation than the scalar one.
In addition, the cross-correlation functions for the scalar GW at different speeds $s_0$ behave differently.
So it is possible to distinguish the tensor polarizations from the scalar ones using PTAs.

Moreover, one notices that when $s_0$ is close to 1 (e.g., the solid black curve), $\zeta(\theta)$ approaches the one predicted in Ref.~\cite{Lee2008ptac} (Fig.~1), which is expected as $s_0$ approaches 1, the scalar GW behaves more like a null wave.
Finally, the cross-correlation functions for the scalar GW are also similar to those in Fig.~6 in Ref.~\cite{Gong:2018cgj} since Fig.~6 was obtained for the special case where the vector polarizations disappear.

\subsection{Gaia mission}\label{sec-gaia}

In order to calculate the astrometric position of a star, one has to imposes the following two conditions \cite{Mihaylov:2018uqm},
\begin{enumerate}
  \item The geodesic of the photon should intersect the one of the Earth at a time, say $t$.
  That is to say, $x^j_\gamma(t)=x^j_\oplus(t)$.
  \item The geodesic of the photon should also intersect the one of the star at the time $t_e+\delta t_e$ with $t_e=t-L$ and $\delta t_e$ of the same order as the perturbations.
      So $x^j_\gamma(t_e+\delta t_e)=x^j_\star(t+\delta t_e)$.
\end{enumerate}
With these two conditions, one finds that
\begin{equation}\label{eq-c-nus}
\begin{split}
  \hat r^j\nu^0+\nu^j=&\Bigg[-\frac{A_{jk}^\text{TT}\hat r^k}{\omega_2L(1+\hat\Omega\cdot\hat r/s_2)}-\frac{\hat\Omega^j/s_2+\hat r_j(2+\hat\Omega\cdot\hat r/s_2)}{2\omega_2L(1+\hat\Omega\cdot\hat r/s_2)^2}\hat r^k\hat r^lA_{kl}^\text{TT}\Bigg]\times\\
  &[\sin\Phi_2(L+t_e)-\sin\Phi_2(t_e)]+\Bigg[-\frac{1}{3}+\Big(\frac{1+\beta}{\alpha}-\frac{3\lambda+2}{2\lambda}s_0^2\Big)\times\\
  &\frac{\hat\Omega\cdot\hat r}{s_0}\Big(2+\frac{\hat\Omega\cdot\hat r}{s_0}\Big)\Bigg]\frac{\hat\Omega^j-\hat\Omega\cdot\hat r\hat r^j}{s_0\omega_0L(1+\hat\Omega\cdot\hat r/s_0)^2}\Theta_0[\sin\Phi_0(L+t_e)-\sin\Phi_0(t_e)].
  \end{split}
\end{equation}
The astrometric position is defined to be $\hat r^j+\delta\hat r^j=-u^\mu_\gamma e_\mu^{\hat j}/f_\oplus$.
After some tedious algebraic manipulations, one obtains
\begin{equation}\label{eq-ch-asmp}
\begin{split}
  \delta\hat r^j(\hat r,\hat\Omega)=&-(\hat r^j\nu^0+\nu^j) +J_0^j(\hat\Omega,\hat r)\Theta(t,0)+J_2^{jkl}(\hat\Omega,\hat r)h_{kl}^\text{TT}(t,0),\\
  \end{split}
\end{equation}
where the first term in the round brackets is given by Eq.~\eqref{eq-c-nus}, and
\begin{gather}
  J_0^j(\hat\Omega,\hat r)=\Bigg\{\frac{1}{3}\frac{\hat\Omega^j}{s_0}-\hat r^j\Big(1+\frac{2}{3}\frac{\hat\Omega\cdot\hat r}{s_0}\Big)+\frac{\hat\Omega\cdot\hat r\hat r^j-\hat\Omega^j}{s_0}\frac{\hat\Omega\cdot\hat r}{s_0}\frac{1+\beta}{\alpha}\nonumber\\
  +\frac{3\lambda+2}{2\lambda}\hat\Omega\cdot\hat r\Big[2\hat\Omega^j+\Big(\frac{\hat\Omega^j}{s_0}-\hat r^j\Big)\hat\Omega\cdot\hat r\Big]\Bigg\}\frac{1}{1+\hat\Omega\cdot\hat r/s_0},\\
  J_2^{jkl}(\hat\Omega,\hat r)=-\frac{\delta^{jl}\hat r^k+\delta^{jk}\hat r^l}{4}+\frac{\hat r^j+\hat\Omega^j/s_2}{2(1+\hat\Omega\cdot\hat r/s_2)}\hat r^k\hat r^j.
\end{gather}
Following Ref.~\cite{Mihaylov:2018uqm}, one assumes the short-wavelength approximation, i.e., $\omega_2L,\,\omega_0L\gg1$.
Then one can drop the first round brackets in Eq.~\eqref{eq-ch-asmp}.
An easy inspection reveals that the resulting expression agrees with the one in Ref.~\cite{Mihaylov:2018uqm} if the speed $s_2$ is set to 1, and the scalar GW is switched off ($\Theta=0$).
The term for the scalar GW, $J_0^j(\hat\Omega,\hat r)\Theta(t,0)$, cannot be rewritten in a form similar to the one for $h_{jk}^\text{TT}$, because Ho\v{r}ava gravity possesses less symmetry which forbids the gauge transformation rendering $g_{00}=-1$.
If the stochastic GW background is still described by Eqs.~\eqref{eq-sgw-scl} and \eqref{eq-tn-ex}, the change in the astrometric position will be
\begin{equation}\label{eq-ch-asmp-sgw}
\begin{split}
  \delta\hat r^j(t,\hat r)=&\int\ud^2\hat\Omega\delta\hat r^j(\hat r,\hat\Omega)\\
  =&\Re\Bigg\{\int_{-\infty}^{\infty}\frac{\ud\omega}{2\pi}\exp(i\omega t)\times\\
  &\int\ud^2\Omega \Big[J_0^j(\hat\Omega,\hat r)\Theta(\omega,\hat\Omega)+\sum_{P=+,\times}J^{jkl}_2(\hat\Omega,\hat r)\epsilon^P_{kl}h_P(\omega,\hat\Omega)\Big]\Bigg\},
  \end{split}
\end{equation}
where the symbol $\Re$ stands for the real part.

The changes in the astrometric positions of two widely separated stars are also correlated as for the frequency shifts \eqref{eq-fs}.
According to Ref.~\cite{Mihaylov:2018uqm}, in order to calculate the correlation, one considers two stars located at directions $\hat r_1=(0,0,1)$ and $\hat r_2=(\sin\theta,0,\cos\theta)$.
For each star, one finds a triad, i.e.,
\begin{equation}\label{eq-trd-1}
  \hat r_1,\quad \hat u^x=(1,0,0),\quad\hat u^y=(0,1,0),
\end{equation}
for the star 1, and
\begin{equation}\label{eq-trd-2}
  \hat r_2,\quad\hat u^\theta=(\cos\theta,0,-\sin\theta),\quad\hat u^\phi=(0,1,0),
\end{equation}
for the star 2.
The correlation $\langle\delta\hat r_1^j(t,\hat r_1)\delta\hat r_2^k(t',\hat r_2)\rangle$ can be factorized, i.e.,
\begin{equation}\label{eq-c-f}
  \langle\delta\hat r_1^j(t,\hat r_1)\delta\hat r_2^k(t',\hat r_2)\rangle=T(t,t')\Gamma_{jk}(\hat r_1,\hat r_2).
\end{equation}
In this expression, $T(t,t')$ is called the temporal correlation factor and is an integral related to the characteristic strains of the GW, as given below,
\begin{gather*}
  T_0(t,t')= \int_{-\infty}^{\infty}\frac{\ud\omega}{4\pi^2}\frac{|\Theta_c(\omega)|}{\omega}[e^{i\omega(t-t')}+e^{-i\omega(t-t')}],\\
  T_2(t,t')=\sum_{P=+,\times}\int_{-\infty}^{\infty}\frac{\ud\omega}{16\pi}\frac{|h^P_c(\omega)|}{\omega}[e^{i\omega(t-t')}+e^{-i\omega(t-t')}].
\end{gather*}
In addition, $\Gamma_{jk}(\hat r_1,\hat r_2)$ is called the spatial correlation factor which contains the information of the polarizations and will be defined below soon.
So it is possible to study the spatial correlation factor separately from the temporal one as it was done in Ref.~\cite{Mihaylov:2018uqm}.

Now, the changes $\delta\hat r_1^j(\hat r_1,\hat\Omega)$ and $\delta\hat r_2^j(\hat r_1,\hat\Omega)$ can be decomposed in the following way
\begin{gather}
\delta\hat r_1^j(\hat r_1,\hat\Omega)=\hat u^j_x\delta\hat r_1^x+\hat u^j_y\delta\hat r_1^y,\\
\delta\hat r_2^j(\hat r_2,\hat\Omega)=\hat u^j_\theta\delta\hat r_2^\theta+\hat u^j_\phi\delta\hat r_2^\phi,
\end{gather}
where $\delta\hat r_1^x=\hat u^x_l\delta\hat r^l_1$, $\delta\hat r_1^y=\hat u^y_l\delta\hat r^l_1$, $\delta\hat r_2^\theta=\hat u^\theta_l\delta\hat r^l_2$ and $\delta\hat r_2^\phi=\hat u^\phi_l\delta\hat r^l_2$ with the index $l$ runs from 1 to 3.
The spatial correlation function for the two stars is
\begin{equation}\label{eq-scf}
\begin{split}
  \Gamma_{jk}=&\int\ud^2\hat\Omega\delta r^j_1(\hat r_1,\hat\Omega)\delta r^k_2(\hat r_2,\hat\Omega)\\
  =&\hat u_j^x\hat u_k^\theta\int\ud^2\hat\Omega\delta r_1^x\delta r_2^\theta+\hat u^y_j\hat u^\phi_k\int\ud^2\hat\Omega\delta r_1^y\delta r_2^\phi\\
  &+\hat u_j^x\hat u_k^\phi\int\ud^2\hat\Omega\delta r_1^x\delta r_2^\phi+\hat u^y_j\hat u^\theta_k\int\ud^2\hat\Omega\delta r_1^y\delta r_2^\theta,
\end{split}
\end{equation}
and it can be shown that the two terms in the last line above vanish \cite{Mihaylov:2018uqm}.
Now, define
\begin{gather}
\Gamma_{x\theta}=\int\ud^2\hat\Omega\delta r_1^x\delta r_2^\theta,\label{eq-sc-xt}\\
\Gamma_{y\phi}=\int\ud^2\hat\Omega\delta r_1^y\delta r_2^\phi.\label{eq-sc-yp}
\end{gather}
In the following, we will calculate these two correlations for the tensor and the scalar GWs with the parameters $\alpha,\,\beta$ and $\lambda$ taking values in Table~\ref{tab-choices-beta-0}.

\begin{figure}
  \centering
  \includegraphics[width=0.45\textwidth]{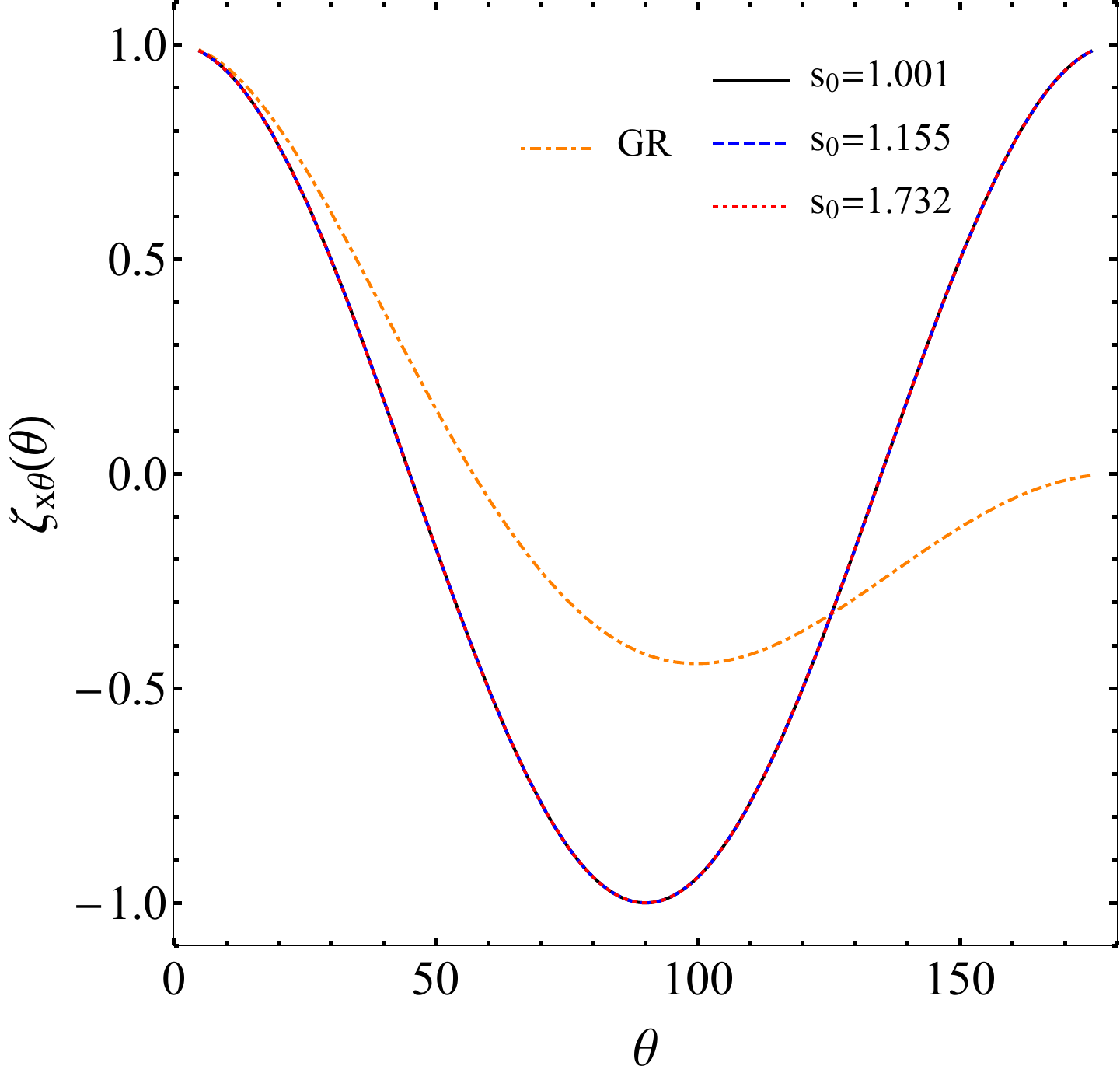}
  \includegraphics[width=0.45\textwidth]{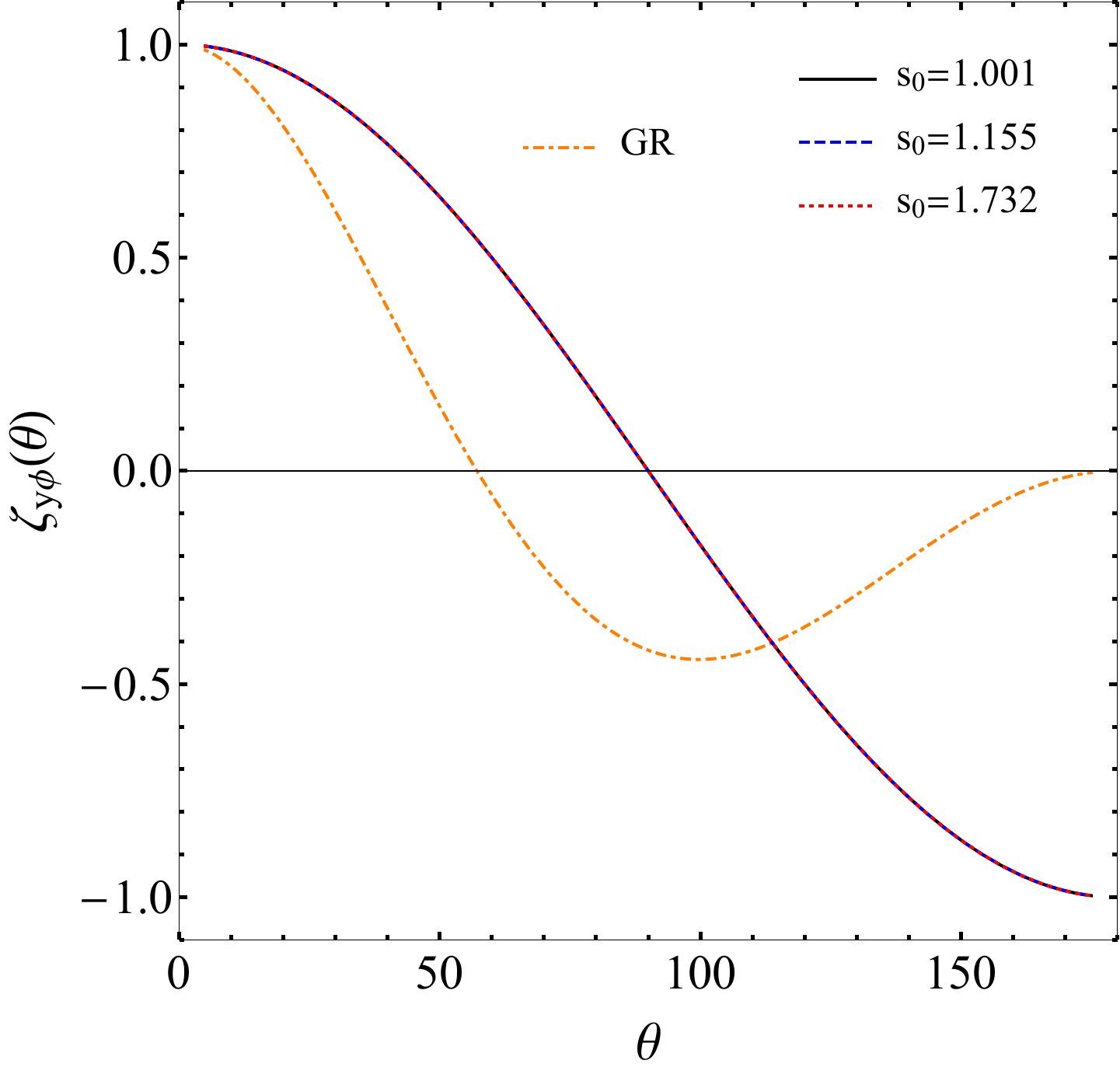}
  \caption{The normalized spatial correlation functions for the tensor and the scalar GWs when the parameters $\alpha,\,\beta$, and $\lambda$ take values in Table~\ref{tab-choices-beta-0}.
  The left panel shows $\zeta_{x\theta}=\Gamma_{x\theta}(\theta)/\Gamma_{x\theta}(0)$, while the right panel shows $\zeta_{y\phi}=\Gamma_{y\phi}(\theta)/\Gamma_{y\phi}(0)$.
  The dot-dashed orange curves in the two panels are for the tensor polarizations plotted using Eq.~\eqref{eq-sc-gr}.
  }\label{fig-sc-beta-0}
\end{figure}

Since $\beta=0$ and $s_2=1$, the spatial correlation functions $\Gamma_{x\theta}$ and $\Gamma_{y\phi}$ take the exact same forms as presented in Ref.~\cite{Mihaylov:2018uqm}.
We will simply quote the results,
\begin{equation}\label{eq-sc-gr}
  \mathcal T(\theta)=\Gamma_{x\theta}^{+,\times}=\Gamma_{y\phi}^{+,\times}=\frac{2\pi}{3}-\frac{14\pi}{3}\sin^2\frac{\theta}{2}-\frac{8\pi\sin^4\frac{\theta}{2}}{\cos^2\frac{\theta}{2}}\ln\sin\frac{\theta}{2},
\end{equation}
which means that the spatial correlation functions are the same for the plus (+) and the cross ($\times$) polarizations.
Now, we will consider the spatial correlation functions due to the scalar GW.
For this purpose, we set $h_{jk}^\text{TT}=0$ in Eq.~\eqref{eq-ch-asmp}.
Because of the complexity of Eq.~\eqref{eq-ch-asmp}, we numerically integrate Eqs.~\eqref{eq-sc-xt} and \eqref{eq-sc-yp}. The results are shown in Fig.~\ref{fig-sc-beta-0}, which actually shows the normalized correlations $\zeta_{x\theta}=\Gamma_{x\theta}(\theta)/\Gamma_{x\theta}(0)$ (the left panel) and $\zeta_{y\phi}=\Gamma_{y\phi}(\theta)/\Gamma_{y\phi}(0)$ (the right panel).
In both panels, the dot-dashed orange curves represent the correlations for the tensor polarizations given by Eq.~\eqref{eq-sc-gr}.
The remaining curves are for the scalar polarizations.
Note that there are actually three curves for the scalar polarizations at different speeds $s_0$ in each panel, but due to the smallness of $\alpha$ and $\lambda$, these curves nearly overlap each other perfectly.
Despite this, it is still possible to distinguish the tensor polarizations from the scalar ones as their correlation functions are rather different.

\subsection{The redshift-astrometric correlation}\label{sec-rac}

One can also form the correlation between the redshift \eqref{eq-fs} and the astrometric position \eqref{eq-ch-asmp}.
For example, a pulsar is at the direction $\hat r_1$ and a star is at $\hat r_2$, then the redshift-astrometric correlation is \cite{Mihaylov:2018uqm}
\begin{equation}\label{eq-red-ast-c}
  \langle z(\hat r_1)\delta r_2^j(\hat r_2)\rangle\propto\Gamma_{z\theta}(\theta)\hat u^j_\theta+\Gamma_{z\phi}(\theta)\hat u^j_\phi,
\end{equation}
with $\Gamma_{z\theta}(\theta)=\int\ud^2\hat\Omega z(\hat r_1)\delta r_2^\theta(\hat r_2)$ and $\Gamma_{z\phi}(\theta)=\int\ud^2\hat\Omega z(\hat r_1)\delta r_2^\phi(\hat r_2)$.
Here, $z(\hat r_1)$ is either Eq.~\eqref{eq-def-i0} or Eq.~\eqref{eq-def-i2}, depending on which polarization one is interested in.
Because of the isotropy of the stochastic GW background, $\Gamma_{z\phi}(\theta)=0$.

\begin{figure}
  \centering
  \includegraphics[width=0.5\textwidth]{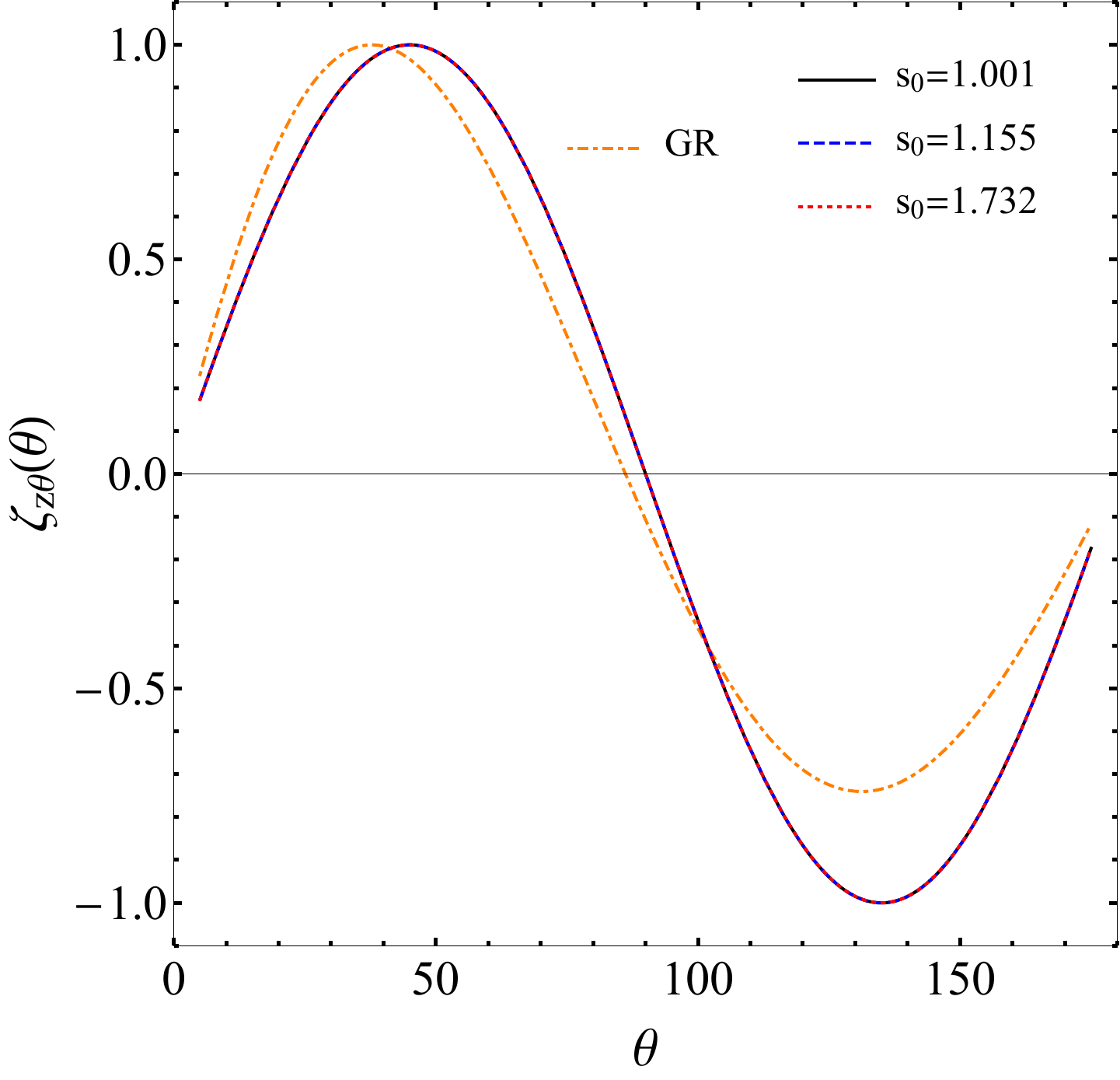}
  \caption{The normalized redshift-astrometric correlation functions $\zeta_{z\theta}(\theta)$ for the tensor polarizations (the dot-dashed orange curve) and the scalar polarizations corresponding to three different speeds given in Table~\ref{tab-choices-beta-0}.
  }\label{fig-cor-zt}
\end{figure}

Ref.~\cite{Mihaylov:2018uqm} considered the redshift-astrometric correlation function for the null tensor GW ($s_2=1$), i.e.,
\begin{equation}\label{eq-redast-gr}
  \Gamma^+_{z\theta}(\theta)=\frac{4\pi}{3}\sin\theta+8\pi\sin^2\frac{\theta}{2}\tan\frac{\theta}{2}\ln\sin\frac{\theta}{2}
\end{equation}
for the plus polarization and $\Gamma^\times_{z\theta}(\theta)=0$ for the cross polarization.
Now, we consider the redshift-astrometric correlation function for the scalar polarization.
For this purpose, set $z(\hat r_1)$ to be Eq.~\eqref{eq-def-i0} and $\delta r_2^\theta=\delta r_2^j\hat u^\theta_j$ with $\delta r_2^j$ given by Eq.~\eqref{eq-ch-asmp} with $h_{jk}^\text{TT}=0$.
Then $\Gamma_{z\theta}(\theta)$ is obtained via the numerical integration and given in Fig.~\ref{fig-cor-zt}, which displays the normalized correlation functions for the tensor polarizations (the dot-dashed orange curve) and the scalar polarizations corresponding to three different speeds listed in Table~\ref{tab-choices-beta-0}.
All correlations are normalized such that the maxima are 1 as it was done in Ref.~\cite{Mihaylov:2018uqm}.
Again there are three curves for the scalar polarization in Fig.~\ref{fig-cor-zt}, but they are close to each other due to the smallness of $\alpha$ and $\lambda$.
As one can see, the redshift-astrometric correlation functions for the tensor and the scalar polarizations are different but the difference is quite limited.
So it would be much easier to distinguish the tensor and the scalar polarizations by measuring the correlation functions $C_0(\theta)$, $C_2(\theta)$ and $\Gamma_{x\theta}$, $\Gamma_{y\phi}$.

\section{Conclusion}\label{sec-con}

In this work, we studied the polarization content of Ho\v rava gravity using the gauge-invariant variable formalism.
The analysis shows that there are the plus, the cross polarizations excited by the transverse-traceless part of the metric perturbation.
There is also the mixture of transverse breathing and the longitudinal polarizations excited by the scalar d.o.f. contained in the theory.
This result is consistent with the fact that Ho\v rava gravity contains three physical d.o.f..
Then, we discussed the experimental constraints on this theory.
In particular, we considered the bounds on the GW speed derived from the observations of GW170817 and GRB 180817A.
It turns out that Ho\v rava gravity is also highly constrained as Einstein-\ae ther theory \cite{Gong:2018cgj}.
Based on these, we discussed whether it is possible to distinguish the tensor and the scalar polarizations using PTAs and the Gaia mission, and thus calculated a variety of cross-correlation functions given in Section~\ref{sec-exps}.
By analyzing the behaviors of the cross-correlation functions, one easily finds out that it might be easy to test the presence of the scalar polarization mode using PTAs and Gaia mission.

\begin{acknowledgements}
This research was supported in part by the Major Program of the National Natural Science Foundation of China under Grant No. 11690021 and the National Natural Science Foundation of China under Grant No. 11475065. E.P acknowledges the hospitality of the School of Physics of the Huazhong University of Science and Technology where part of this work was carried out.
S. H. was also supported by China Postdoctoral Science Foundation (No. 2018M632822).
\end{acknowledgements}


%

\end{document}